\documentclass[journal]{IEEEtran}

\usepackage{cite}

\ifCLASSINFOpdf
   \usepackage[pdftex]{graphicx}
   \graphicspath{{../pdf/}{../jpeg/}}
   \DeclareGraphicsExtensions{.pdf,.jpeg,.png}
\else
\fi

\usepackage{amsmath}
\usepackage{array}
\usepackage{standalone}
\usepackage{etoolbox}
\newtoggle{arXiv}%\toggletrue{arXiv}
\newtoggle{bibrun}%\toggletrue{bibrun}					
\usepackage{newtxtext, newtxmath}
\usepackage{graphicx}
\usepackage{color, calc}
\usepackage{array, booktabs}
\newcolumntype{L}{>{$}l<{$}}
\newcolumntype{C}{>{$}c<{$}}
\newcolumntype{R}{>{$}r<{$}}
\usepackage{tikz, tikz-3dplot}
\usetikzlibrary{arrows, arrows.meta, calc, positioning}
\usetikzlibrary{decorations.pathmorphing}	
\tikzset{>=Stealth}
\usepackage{pgfplots}
\usepackage{siunitx, physics}
\usepackage{enumitem}
\setlist[description]{labelindent=0pt, leftmargin=\parindent, font=\normalfont\itshape}
\usepackage{url}
\usepackage{subfigure}
\usepackage{textcomp}
\usepackage{eurosym}

\begin{document}
\title{Simultaneous imaging of hard and soft biological tissues in a low-field dental MRI scanner}

\author{\IEEEauthorblockN{
		Jos\'e~M.~Algar\'{\i}n\IEEEauthorrefmark{1},
		Elena~D\'{\i}az-Caballero\IEEEauthorrefmark{2},
		Jose~Borreguero\IEEEauthorrefmark{1},
		Fernando~Galve\IEEEauthorrefmark{1},
		Daniel~Grau-Ruiz\IEEEauthorrefmark{2},
		Juan~P.~Rigla\IEEEauthorrefmark{2},
		Rub\'en~Bosch\IEEEauthorrefmark{1},
		Jos\'e~M.~Gonz\'alez\IEEEauthorrefmark{2},
		Eduardo~Pall\'as\IEEEauthorrefmark{1},
		Miguel~Corber\'an\IEEEauthorrefmark{1},
		Carlos~Gramage\IEEEauthorrefmark{1},
		Santiago~Aja-Fern\'andez\IEEEauthorrefmark{3},
		Alfonso~Ríos\IEEEauthorrefmark{2},
		Jos\'e~M.~Benlloch\IEEEauthorrefmark{1}, and
		Joseba~Alonso\IEEEauthorrefmark{1}}
	
	\IEEEauthorblockA{\IEEEauthorrefmark{1}MRILab, Institute for Molecular Imaging and Instrumentation (i3M), Spanish National Research Council (CSIC) and Universitat Polit\`ecnica de Val\`encia (UPV), 46022 Valencia, Spain}\\
	\IEEEauthorblockA{\IEEEauthorrefmark{2}Tesoro Imaging S.L., 46022 Valencia, Spain}\\
	\IEEEauthorblockA{\IEEEauthorrefmark{3}Laboratorio de Procesado de Imagen, Universidad de Valladolid, 47011 Valladolid, Spain}\\%
	
\thanks{Corresponding author: J. Alonso (joseba.alonso@i3m.upv.es).}}

\markboth{Journal of \LaTeX\ Class Files,~Vol.~14, No.~8, August~2015}%
{Shell \MakeLowercase{\textit{et al.}}: Bare Demo of IEEEtran.cls for IEEE Journals}

\maketitle

\begin{abstract}
Magnetic Resonance Imaging (MRI) of hard biological tissues is challenging due to the fleeting lifetime and low strength of their response to resonant stimuli, especially at low magnetic fields. Consequently, the impact of MRI on some medical applications, such as dentistry, continues to be limited. Here, we present three-dimensional reconstructions of \emph{ex-vivo} human teeth, as well as a rabbit head and part of a cow femur, all obtained at a field strength of only 260~mT. These images are the first featuring soft and hard tissues simultaneously at sub-Tesla fields, and they have been acquired in a home-made, special-purpose, pre-medical MRI scanner designed with the goal of demonstrating dental imaging at low field settings. We encode spatial information with two variations of zero-echo time (ZTE) pulse sequences: Pointwise-Encoding Time reduction with Radial Acquisition (PETRA) and a new sequence we have called Double Radial Non-Stop Spin Echo (DRaNSSE), which we find to perform better than the former. For image reconstruction we employ Algebraic Reconstruction Techniques (ART) as well as standard Fourier methods. A noise analysis of the resulting images shows that ART reconstructions exhibit a higher signal to noise ratio with a more homogeneous noise distribution.
\end{abstract}

 \ifCLASSOPTIONpeerreview
 \begin{center} \bfseries EDICS Category: 3-BBND \end{center}
 \fi
\IEEEpeerreviewmaketitle

\section{Introduction}
%I know that the introduction will suffer a lot of changes. But let me suggest something that should not change. This is the take-home message of the first four paragraphs. I think that it will help to write the introduction a lot.
%
%\begin{enumerate}
%	\item Why imaging of hard tissues is important?
%	\item Where is the state of the art for hard tissue imaging?
%	\item What are the current problems of the state of the art?
%	\item What do we propose to overcome the described issues in previous paragraphs?
%\end{enumerate}

\IEEEPARstart{M}{agnetic} Resonance Imaging (MRI, \cite{BkHaacke}) plays an indispensable role in healthcare. In particular, its performance is unrivaled for soft tissues, being the only known technique capable of \emph{in-vivo} imaging of deep biological tissues with high spatial resolution and tissue contrast while avoiding harmful ionizing radiation \cite{Bercovich2018}. Despite MRI's unquestionable success, imaging hard tissues (such as bone, tendons, dentin or enamel) remains problematic \cite{Mastrogiacomo2019}. This is due to the fleeting lifetime and low strength of signals emitted by solids, as opposed to the case of soft and non-solid tissues \cite{BkDuer}. In the latter, the deleterious dipole-dipole magnetic interaction between neighboring spins averages out much faster than any significant timescale in imaging processes. This results in strong, long-lived signals which are routinely exploited for the high quality images typical for MRI. In hard tissues, however, spins are to a good approximation fixed with respect to one another, and the dipole-dipole pattern vanishes exclusively at the so-called ``magic angle'' \cite{Oatridge2001}. This means that every spin is subject to a noisy environment created by the surrounding spins. Since this contribution does not average out, the MR signal coherence of hard biological tissues is typically lost in hundreds of microseconds \cite{Funduk1984,Schreiner1991}.

The above background justifies the rather meager penetration of MRI in dentistry \cite{Niraj2016}. Instead, odontologists heavily use X-ray based technologies \cite{Shah2014}. However, these come together with a number of detrimental aspects, such as the use of ionizing radiation, low soft tissue contrast, and unreliable revelation of pulp diseases \cite{Newton2009} or tooth cracks and fractures \cite{Brady2014,Idiyatullin2016}. All these inconveniences can be overcome with MRI \cite{Idiyatullin2011,Idiyatullin2016}, albeit with its own particular challenges. Aside from the aforementioned technical difficulties in detecting hard biological tissues, MRI scanners are typically expensive to acquire, site, operate and maintain, mostly due to the high magnetic fields involved \cite{Marques2019}. Low-field systems constitute a promising inexpensive alternative to standard MRI setups \cite{Sarracanie2015,Marques2019}, but the reduced signal-to-noise ratio (SNR) can easily lead to long acquisition times incompatible with clinical conditions.

%******************************************************************
\begin{figure*}
	\centering
	\includegraphics[width= 2\columnwidth]{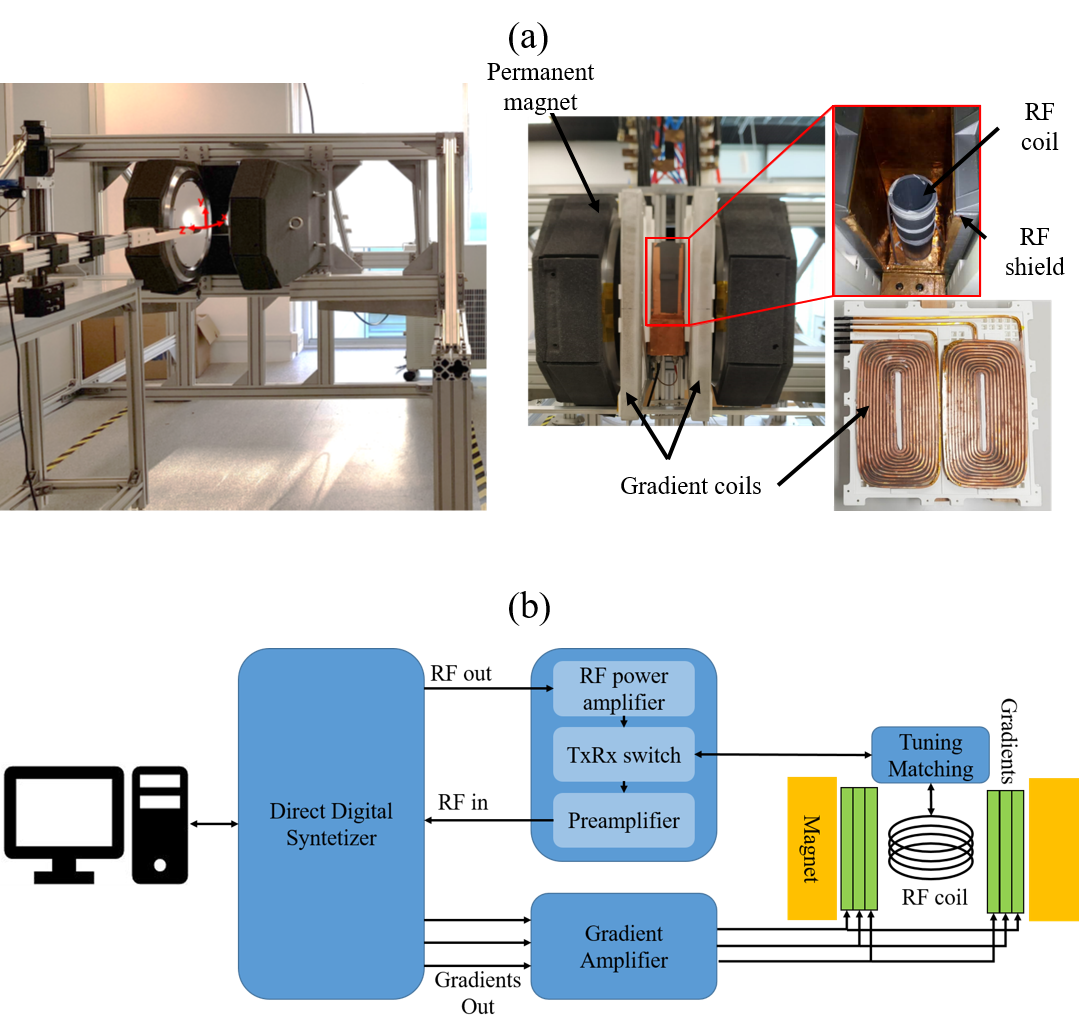}
	\caption{(a) (Left) Photograph of the main magnet installed on the support structure. (Right) Photographs of ``DentMRI - Gen I'', showing a general overview of the magnet and the components used for building the gradient and radio-frequency systems. (b) Sketch of the ``DentMRI - Gen I'' scanner.}
	\label{fig:setup}
\end{figure*}
%******************************************************************

In this article we present ``DentMRI - Gen I'' (Fig.~\ref{fig:setup}), a home-made special-purpose MRI scanner designed with the goal of demonstrating dental imaging at low field settings (Sec.~\ref{sec:scanner}). In Sec.~\ref{sec:seqs} we describe two variations of zero-echo time (ZTE) pulse sequences: Pointwise-Encoding Time reduction with Radial Acquisition (PETRA, \cite{Weiger2012,Grodzki2012}); and Double Radial Non-Stop Spin Echo (DRaNSSE), which we have devised to address sampling and contrast limitations we encounter with PETRA. We discuss in Section~\ref{sec:reconst} two different mathematical treatments for image reconstruction, and finally demonstrate the capability of ``DentMRI - Gen I'' to produce high quality combined images of soft and hard biological tissues at $\approx$260~mT (Sec.~\ref{sec:images}). These constitute, to our best knowledge, the first images featuring soft and hard tissues simultaneously at sub-Tesla fields. We reconstruct images of a rabbit head, bare human teeth and a cow femur with PETRA and DRaNSSE, using both Algebraic Reconstruction Techniques (ART, \cite{Karczmarz1937,Gordon1970,Gower2015}) and traditional Fourier transformation. We find that DRaNSSE can perform significantly better than PETRA and that ART results in higher quality reconstructions than Fourier spectral analysis with our settings.

%******************************************************************
%******************************************************************
%******************************************************************

\section{``DentMRI - Gen I'' Scanner}
\label{sec:scanner}

``DentMRI - Gen I'' (Fig.~\ref{fig:setup}(a)) is our first-generation MRI dental scanner, designed and built to demonstrate hard tissue imaging techniques in low magnetic field settings. Operating at a field of $\approx$260~mT provided by a permanent magnet, all parts are standard commodities, 3D-printable, or inexpensive to machine. The total cost of all components adds up to $\approx$150~k\euro, a small fraction of the price of high field scanners (based on super-conducting magnets) previously used for dental imaging \cite{Idiyatullin2011,Weiger2012,Ludwig2016,Idiyatullin2016,Niraj2016}.

``Gen I'' is meant for technology-demonstration purposes rather than for immediate clinical validation. We have therefore restricted our use to \emph{ex-vivo} image acquisitions of samples accommodated in a cylindrical field of view of height $\approx$100~mm and diameter $\approx$45~mm, even if the gap between magnet poles is significantly larger. We also enclose the samples inside a Faraday cage to isolate our single radio-frequency (rf) coil from the magnetic gradient coils and electromagnetic interference noise present in the laboratory, which is otherwise unshielded. In the following we describe the main hardware components of ``Gen I'', and leave for Section~\ref{sec:concl} a discussion on possible approaches towards \emph{in-vivo} combined hard and soft tissue imaging with a prospective ``DentMRI - Gen II''.

%******************************************************************
%******************************************************************
%******************************************************************

\subsection{Main magnet}
\label{sec:magnet}

In MRI systems the quantization axis, spin resonance (Larmor) frequency and, to a large extent, the signal-to-noise ratio of the detected signals are typically determined by a strong homogeneous magnetic field \cite{BkHaacke}. Here we employ a ``C''-shaped permanent NdFeB magnet (Sabr Enterprises LLC, Fig.~\ref{fig:setup}(a)), providing a main evolution field of $\approx$260~mT. The field is shimmed down to an homogeneity of $\approx$20~parts-per-million (ppm) over a spherical region of 150~mm in diameter. The magnet is the heaviest component of the scanner with a mass of around 940~kg for a distance between magnet poles of $\approx$220~mm. In order to distribute its weight over a $\approx$5~m$^2$ floor surface (compliant with typical architectural regulations), we designed and constructed a support structure consisting mostly of profiles manufactured out of Aluminum 6063. Although this was specified to be non-magnetic, it induces field-strength inhomogeneities at the 4000~ppm level. These are mostly linear and we regularly shim down to $\approx$4~ppm during operation with the magnetic gradient system (Sec.~\ref{sec:grads}).

%******************************************************************
%******************************************************************
%******************************************************************

\subsection{Magnetic gradient system}
\label{sec:grads}

Linear magnetic field inhomogeneities (magnetic gradients) spatially encode information in MRI setups \cite{BkHaacke}. Intense gradient fields lead to efficient encoding, i.e. strong resolving power. ``Gen I'' is equipped with a gradient system capable of reaching strengths $>400$~mT/m along any spatial direction, sufficient for sub-mm spatial resolution.

The gradient system consists of three pairs of planar coils. Each pair produces a magnetic field pointing in the same direction as the main field, and a gradient in magnitude along a Cartesian axis in the laboratory frame of reference. Each of the $z$ ($x,y$) coils is formed by one (two) lobes (see Fig.~\ref{fig:setup}(a)). We fabricated the coils with hollow-tube Oxygen-Free-High-Conductivity (OFHC) copper, with low ohmic losses ($<\SI{50}{m \ohm}$ per pair) and the possibility of heat removal by running cooling water through the inner conduct. We can stably operate at 100~\% duty cycle with a total power consumption $<8$~kW. In order to pulse the magnetic gradient fields in short times of 50-\SI{100}{\micro s}, we wound each coil into only a few loops for a total pair inductance $<\SI{100}{\micro H}$ in all three cases. The size and disposition of the coils ensures deviations $<10$~\% from perfect linearity over a spherical region of 10~cm in diameter. The space between loops is filled with translux D150 epoxy to avoid shorts and improve mechanical stability. We drive the coils with bipolar amplifiers from International Electric Co. (GPA-400-750), which can ramp from 0 to $\pm$400~A in \SI{100}{\micro s} with our loads.

%******************************************************************
%******************************************************************
%******************************************************************

\subsection{Radio-frequency system}
\label{sec:rf}

In MRI, rf electronics are required to excite the sample spins and detect their response for amplification, digitization and, ultimately, data processing as required for image reconstruction \cite{BkHaacke}. We employ a single coil for both sample excitation and signal detection. The solenoid, of length $\approx$100~mm, $\approx$52~mm in diameter and with 20 windings with copper wire of $\approx$0.4~mm diameter, is shown in Fig.~\ref{fig:setup}(a). The solenoid includes a gap capacitor to homogenize the current along the coil wire, resonant at $\omega_\text{c}\approx 2\pi\cdot$11~MHz with a quality factor $Q \approx 32$. We keep the $Q$ intentionally low to achieve excitation and detection bandwidths $>1$~MHz, compatible with high spatial resolution images. The rf electronics for fine tuning and impedance matching of the resonant circuit are placed next to the solenoid (Fig.~\ref{fig:setup}(b)). Both the coil and matching electronics are enclosed in conducting Faraday boxes to avoid the deleterious effect of electromagnetic noise in the laboratory and coupling to the gradient coils. We 3D-printed the housing structure out of polylactic acid for mechanical stability (Fig.~\ref{fig:setup}(a)). 

In transmission (Tx) mode, we drive the solenoid from an rf power amplifier (RFPA-4/11-2000 from Barthel HF-Technik GmbH) fed by a direct digital synthesizer on our field-programmable-gate-array (FPGA) board (RadioProcessor-G from SpinCore Technologies Inc, see Sec.~\ref{sec:CSGUI}). The RFPA output is low-pass filtered and sent to the solenoid coil (probe) for sample excitation. This scheme is shown in Fig.~\ref{fig:setup}(b).

A TxRx switch (Barthel HF-Technik GmbH, dead time $<\SI{5}{\micro s}$ at 11~MHz) commutes the system operation from transmission to reception (Rx). A low noise amplifier (Barthel HF-Technik GmbH, gain 39~dB and noise factor 1.0~dB) amplifies the analog signal induced by the precessing protons on the coil. After low-pass filtering, we digitize and digitally down-convert and filter the signal directly on the FPGA board.

%******************************************************************
%******************************************************************
%******************************************************************

\subsection{Control system and Graphical User Interface}
\label{sec:CSGUI}

The FPGA board is the main component in the experimental control system. This is plugged into a Peripheral Component Interconnect (PCI) slot on the motherboard of a control computer, and allows us to: i) generate low power rf pulses which are amplified for coherent spin excitation; ii) generate three independent low-frequency and low-power outputs which are amplified and fed to the gradient coils for spatial information encoding; iii) read in, digitize, down-convert and filter the MR signal emitted by the sample; iv) execute all of the above operations synchronously; and v) communicate bidirectionally  with the control computer.

We have programmed our own Graphical User Interface (GUI) in Matlab, where we design pulse sequences, set the individual pulse parameters independently, configure data acquisition and filtering settings, and visualize the received data and reconstructed objects. Additionally, we have written an intermediate layer in C/C++ to interact from the GUI with the provided drivers for the RadioProcessor-G.

To perform image reconstruction we use a PC with an Intel Core i7-7700 CPU (6.65~GHz, 4 main processors, 8 logic processors), an NVIDIA GeForce GTX 750 Ti GPU (640 CUDA cores, 1.4 TFLOPS, 2GB memory) and 32 GB RAM.

%******************************************************************
%******************************************************************
%******************************************************************

\section{Pulse sequences}
\label{sec:seqs}

%******************************************************************
\begin{figure}
	\includegraphics[width= 0.9\columnwidth]{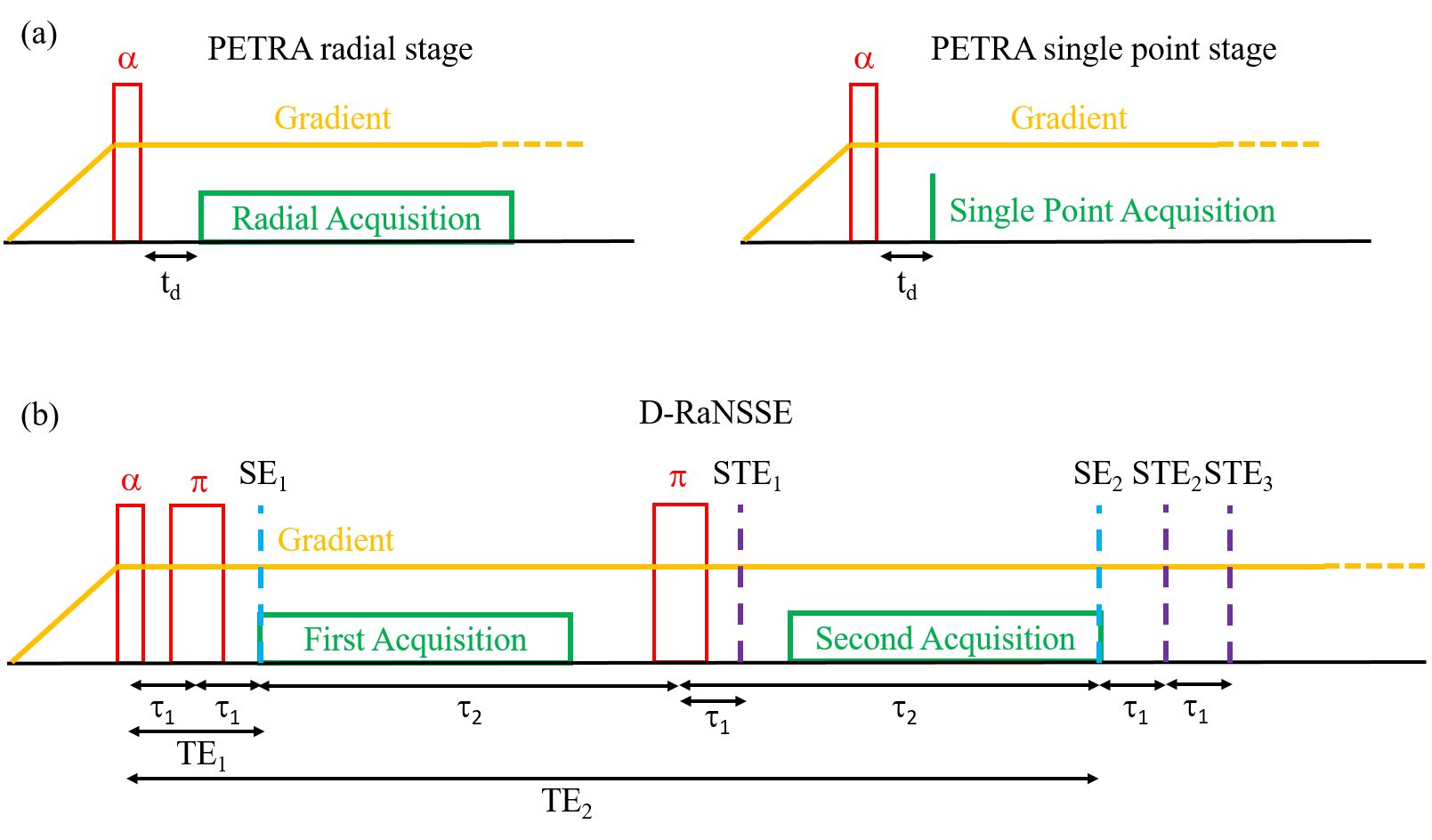}
	\caption{Sequence diagram for a single repetition of (a) PETRA and (b) DRaNSSE.}
	\label{fig:seqZTE}
\end{figure}
%******************************************************************

MRI pulse sequences are designed to manipulate the sample magnetization: resonant rf pulses lead to coherent rotations, and magnetic-field-gradient (or simply gradient) pulses encode spatial information. The short lifetime of hard tissue signals imposes the use of special-purpose pulse sequences. These typically force a high-intensity constructive interference signal (echo) over very short timescales, or even simultaneously excite the sample and detect its response \cite{Mastrogiacomo2019}. Ultra-short echo time (UTE, \cite{Tyler2007}) and zero-echo time (ZTE, \cite{Weiger2012}) pulse sequences are prominent examples of the former approach, and SWeep Imaging with Fourier Transformation (SWIFT, \cite{Idiyatullin2011}) of the latter.

ZTE and SWIFT are better suited than UTE for ultra-short $T_2$ tissues such as those present in teeth. If the scanner hardware allows for rf excitations with sufficient bandwidth, ZTE is the most convenient choice \cite{Weiger2011}. Since the ``DentMRI - Gen I'' scanner has a rather small field of view and we have a fast 2~kW rf power amplifier, we acquired all images below with two variations of standard ZTE pulse sequences (Fig.~\ref{fig:seqZTE}): Pointwise Encoding Time Reduction with Radial Acquisition (PETRA, \cite{Grodzki2012}); and a sequence we have devised to overcome specific $k$-space (spatial frequency space) coverage and contrast limitations, Double Radial Non-Stop Spin Echo (DRaNSSE).

In PETRA (Fig.\ref{fig:seqZTE}(a)), a hard rf pulse homogeneously excites the sample after the onset of the encoding gradient fields. Data acquisition starts next, after a dead time usually set by the TxRx switch and the rf coil ring-down (Sec.~\ref{sec:rf}). Every acquisition follows a radial direction (spoke) in a 3-dimensional $k$-space. The center of $k$-space is not sampled in this way due to the finite dead time, so it is filled in a pointwise manner on a Cartesian grid. Once all radial spokes have been sampled, object reconstruction can be carried out either by inverse discrete Fourier transformation (which requires regridding and interpolation operations), or with Algebraic Reconstruction Techniques (ART, see Sec.~\ref{sec:reconst}, \cite{Karczmarz1937,Gordon1970,Gower2015}).

PETRA suffers from two main constraints. On the one hand, scan times can become exceedingly long when the pointwise-filled gap is large. This is accentuated at low fields, due to the slow ring-down of the Tx coil, which happens with a time constant $\tau=Q/\omega_\text{c}$. On the other hand, it is not possible to have $T_2$ contrast with PETRA, since the echo time (TE) is defined to be zero. Our solution to both problems is our new sequence DRaNSSE (Fig.\ref{fig:seqZTE}(b)), also inspired on ZTE. When we do acquire with PETRA, we can subtract two independent scans with different parameters for $T_2$ contrast, following a previously existing scheme with UTE pulse sequences \cite{Rahmer2007}. In one acquisition dead times are short, so hard tissues still appear bright, while in the other we use intentionally long dead times, to ensure hard-tissue contributions fade away before readout.

In DRaNSSE, after the initial rf pulse, two refocusing $\pi$-pulses produce two different spin echoes. The first one (SE$_1$) is induced at an echo time (TE$_1$) as short as possible to include contributions from both hard and soft tissues. The second echo (SE$_2$) occurs at a later time (TE$_2$), when the short-lived signal from hard tissues has already faded away. A first acquisition takes place between SE$_1$ and the second refocusing pulse, corresponding to radial spokes in $k$-space from $k = 0$ to the maximum sampled value ($k_\text{max}$); the second acquisition starts after the second $\pi$-pulse and ends at TE$_2$, sampling from $-k_\text{max}$ to $k = 0$. The second acquisition could go beyond SE$_2$, allowing for higher quality imaging of soft tissues. However, the occurrence of multiple unwanted stimulated echoes due to small experimental imperfections (STE$_1$-STE$_3$ in Figure~\ref{fig:seqZTE}(b)) can degrade image quality. STE$_1$ appears a time $\tau_1 = \text{TE}_1 / 2$ after the second $\pi$-pulse, while STE$_2$ and STE$_3$ occur at times $\tau_1$ and $2\tau_1$ after SE$_2$, respectively. Since $\tau_1$ in our sequences is much shorter than $\tau_2$ (the time between the start of the first acquisition and the center of the second refocusing pulse), we find it convenient to acquire the signal before SE$_2$ to avoid reconstruction artifacts.

\section{Fourier and Algebraic Reconstruction Techniques}
\label{sec:reconst}

Spatial encoding in MRI relies on inhomogeneous magnetic fields, which provide a Larmor or spin-precession frequency ($\omega_\text{L}$) dependent on the position of the nuclei in the Region of Interest (RoI). Mathematically, this can be expressed as
\begin{equation}
\omega_\text{L}(\vec{r},t) = \gamma\left|\vec{B}(\vec{r},t)\right|,
\end{equation}
where $\gamma$ is the gyromagnetic ratio ($\approx 2\pi\cdot 42$~MHz/T for protons) and $\vec{B}$ is the magnetic field at position $\vec{r}$ at time $t$. As the pulse sequence evolves, the phase acquired by the spins depends on their position:
\begin{equation}\label{eq:phase}
\Phi(\vec{r},t)=\int_0^t \omega_\text{L}(\vec{r},t') \text{d}t'=\int_0^t \gamma |\vec{B}(\vec{r},t')| \text{d}t'.
\end{equation}
During their precession, spins induce a time-varying signal with the interference of all spins on a nearby detector:
\begin{equation}\label{eq:EF}
s(t)\propto \int_\text{RoI} \text{e}^{-\text{i}\Phi(\vec{r},t)}\rho(\vec{r}) \text{d}\vec{r},
\end{equation}
where $\rho(\vec{r})$ is the spin density distribution in the RoI. This signal is then digitized during a readout or acquisition window, and we apply one of the following mathematical tools to reconstruct an image.

\subsection{Discrete Fourier Transform reconstruction}

Typically, scanners make use of linear gradient fields, in the presence of which the Larmor frequency varies also linearly with position. In this scenario, an inverse Fourier Transformation (FT) of $s(t)$ suffices to reconstruct $\rho(\vec{r})$, since the integral in Eq.~(\ref{eq:phase}) after down-mixing is trivial and Eq.~(\ref{eq:EF}) becomes
\begin{equation}\label{eq:FT}
s(t)=\int \text{e}^{-\text{i}\gamma G_z z t}\rho(z) \text{d}z.
\end{equation}
Here we have assumed, without loss of generality, that a gradient of strength $G_z$ points along the $z$-axis. Since the detected signal is discretized in time by an analog-to-digital converter, fast Fourier transform protocols can be used to map a discrete $k$-space onto a discrete reconstruction in real space \cite{BkHaacke}.

Discrete Fourier transforms are computationally efficient and applicable in many useful scenarios, but this simple approach may yield suboptimal results when the sampling data does not adjust to a Cartesian grid in $k$-space. This occurs, for instance, if gradients are time dependent, discretization times are not homogeneously distributed or, as in the present work, $k$-space is sampled radially or following curved trajectories. In such cases, the acquired data must be pre-processed (e.g. with regridding, density compensation or interpolation operations) prior to Fourier transformation \cite{Fessler2007}. To perform Fourier reconstruction we first interpolate the $k$-space data to a Cartesian grid and then apply a Fast Fourier Transform protocol to the interpolated data.

\subsection{Reconstruction based on an Encoding Matrix}

An alternative to pre-processing and Fourier operations in non-Cartesian $k$-space sampling is to build a linear forward model for the system response to the applied pulse sequence, define a cost function for the reconstruction, optionally add regularization terms to penalize unrealistic results, and solve a linear inversion problem \cite{Fessler2010}.

Once the time-dependent signal resulting from the interference of the precessing nuclei has been recorded and discretized, $s(t)$ becomes a vector $\mathbf{S}$ of length equal to the number of time steps $n_\text{t}$, $\rho(\vec{r})$ becomes a vector $\boldsymbol{\rho}$ of length equal to the number of voxels $n_\text{v}$, and $\exp{-i\Phi(\vec{r},t)}$ becomes the Encoding Matrix $\mathbf{\Phi}$ with $n_\text{t}$ rows and $n_\text{v}$ columns. Equation~(\ref{eq:EF}) thus changes to
\begin{equation}\label{eq:EM}
\mathbf{S}=\mathbf{\Phi}\boldsymbol{\rho},
\end{equation}
and $\boldsymbol{\rho}$ can be obtained by direct inversion of the Encoding Matrix as $\mathbf{\Phi}^{-1}\mathbf{S}$, or by any other means of solving the system of linear equations, e.g. by iterative algorithms such as Algebraic Reconstruction Techniques (ART) \cite{Karczmarz1937,Gordon1970,Gower2015}. ART estimates $\rho(\vec{r})$ based on the recursive equation
\begin{equation}\label{eq:ART}
\boldsymbol{\rho}_n = \boldsymbol{\rho}_{n-1}+\lambda\frac{S_i-\mathbf{\Phi}_i\cdot\boldsymbol{\rho}_{n-1}}{\left\|\mathbf{\Phi}_i\right\|}\mathbf{\Phi}_i^*,
\end{equation}
where $\lambda$ is a control parameter, $S_i$ is the $i^\text{th}$ component in vector $\mathbf{S}$, $\mathbf{\Phi}_i$ is the $i^\text{th}$ row in the encoding matrix $\mathbf{\Phi}$, and $\boldsymbol{\rho}_0$ can be set to zero. The estimated solution $\boldsymbol{\rho}_n$ is updated $n_\text{t}\cdot n_\text{it}$ times through Eq.~(\ref{eq:ART}), where $n_\text{it}$ stands for the overall number of ART iterations.

Although iterative methods such as ART are computationally slow compared to discrete Fourier transforms, we find in this work that they can vastly outperform Fourier reconstruction in other relevant metrics.

\section{Hard and soft tissue images}
\label{sec:images}

In the following, we will first provide images in which hard and soft biological tissues are simultaneously visible, we will then show 3D reconstructions of bare human teeth, and we will end the section with quantitative comparisons between the performance of PETRA and DRaNSSE pulse sequences, and between Fourier and ART-based mathematical reconstructions. Table~\ref{tab:imageParameters} contains the parameters used for all images in this section.

\begin{table*}[]
	\caption{Image acquisition parameters. ``NA'' stands for ``not applicable''.}
	\label{tab:imageParameters}
\begin{tabular}{|c|c|c|c|c|c|c|c|c|c|c|c|c|}
\hline
Image        & Sequence & \begin{tabular}[c]{@{}c@{}}Flip\\ angle ($^\circ$)\end{tabular} & \begin{tabular}[c]{@{}c@{}}Pulse \\ time (us)\end{tabular} & \begin{tabular}[c]{@{}c@{}}FOV\\ (mm$^3$)\end{tabular} & \begin{tabular}[c]{@{}c@{}}Pixel \\ size\\ (mm)\end{tabular} & \begin{tabular}[c]{@{}c@{}}Dead\\ time (us) /\\ Acquisition\\ time (us)\end{tabular} & TE (us) & \begin{tabular}[c]{@{}c@{}}TR\\ (ms)\end{tabular} & \begin{tabular}[c]{@{}c@{}}Radial\\ spokes \end{tabular} & \begin{tabular}[c]{@{}c@{}}Single\\ points\end{tabular} & Scans & \begin{tabular}[c]{@{}c@{}}Scan\\ time\\ (min)\end{tabular} \\ \hline
I. Fig. 3    & PETRA    & 58                                                       & 10                                                         & 46$\times$54$\times$42                              & 0.5                                                          & 90 / 710                                                                             & NA      & 10                                                & 4098                                                   & 848                                                     & 75    & 61                                                          \\ \hline
II. Fig. 3   & PETRA    & 58                                                       & 10                                                         & 46$\times$54$\times$42                              & 0.5                                                          & 1,000 / 4,000                                                                        & NA      & 10                                                & 4098                                                   & 3904                                                    & 25    & 33                                                          \\ \hline
III. Fig. 4  & PETRA    & 90                                                       & 10                                                         & 42$\times$45$\times$38                              & 0.5                                                          & 200 /400                                                                             & NA      & 20                                                & 3130                                                   & 7624                                                    & 150   & 540                                                         \\ \hline
IV. Fig. 5   & PETRA    & 90                                                       & 9.1                                                        & 46$\times$48$\times$32                              & 1                                                            & 85 / 915                                                                             & NA      & 50                                                & 4896                                                   & 40                                                      & 7     & 29                                                          \\ \hline
V. Fig. 5    & PETRA    & 90                                                       & 9.1                                                        & 46$\times$48$\times$32                              & 1                                                            & 1,000 / 2,000                                                                        & NA      & 50                                                & 4896                                                   & 1528                                                    & 7     & 37                                                          \\ \hline
VI. Fig. 5   & DRaNSSE  & 90                                                       & 9.1                                                        & 46$\times$48$\times$32                              & 1                                                            & NA / 1,000                                                                           & 60      & 50                                                & 4896                                                   & NA                                                      & 16    & 65                                                          \\ \hline
VII. Fig. 5  & DRaNSSE  & 90                                                       & 9.1                                                        & 46$\times$48$\times$32                              & 1                                                            & NA / 1,000                                                                           & 10,000  & 50                                                & 4896                                                   & NA                                                      & 16    & 65                                                          \\ \hline
VIII. Fig. 6 & DRaNSSE  & 90                                                       & 10                                                         & 44$\times$52$\times$42                              & 1                                                            & NA / 1,000                                                                           & 60      & 50                                                & 1426                                                   & NA                                                      & 26    & 31                                                          \\ \hline
IX. Fig. 6   & DRaNSSE  & 90                                                       & 10                                                         & 44$\times$52$\times$42                              & 1                                                            & NA / 1,000                                                                           & 10,000  & 50                                                & 1426                                                   & NA                                                      & 26    & 31                                                          \\ \hline
X. Fig. 6    & PETRA    & 90                                                       & 10                                                         & 44$\times$52$\times$42                              & 1                                                            & 90 / 910                                                                             & NA      & 50                                                & 1426                                                   & 64                                                      & 12    & 15                                                          \\ \hline
XI. Fig. 6   & PETRA    & 90                                                       & 10                                                         & 44$\times$52$\times$42                              & 1                                                            & 1,000 / 4,000                                                                        & NA      & 50                                                & 1426                                                   & 496                                                     & 9     & 15                                                          \\ \hline
\end{tabular}
\end{table*}

\subsection{Combined tissue imaging: rabbit head}

\begin{figure*}
	\includegraphics[width= 2\columnwidth]{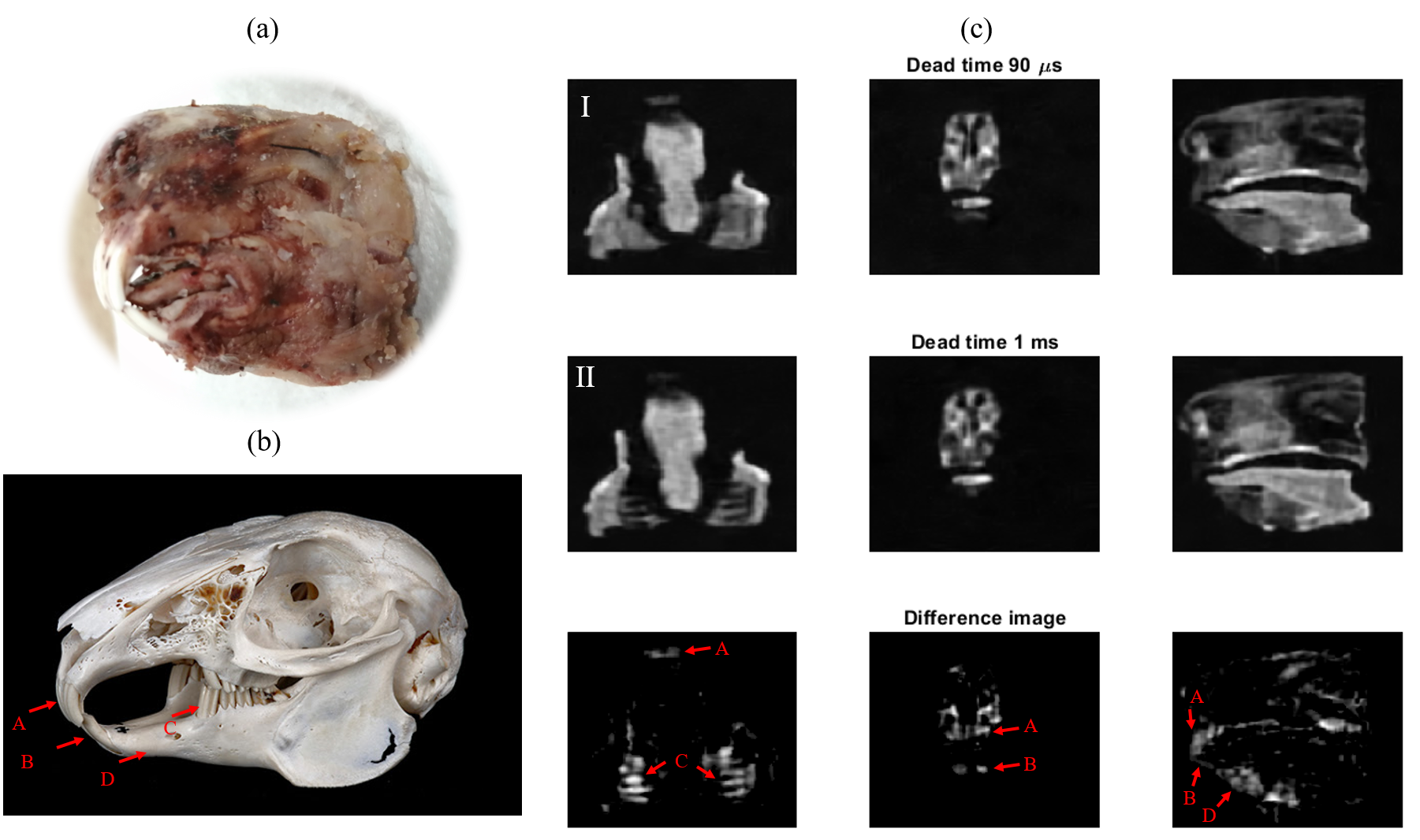}
	\caption{(a) Picture of the scanned rabbit head. (b) Picture of a rabbit skull, taken from Gabrielle Ochnik, Pinterest. (c) Top: Single slices for \SI{90}{\micro s} dead time; middle: the same slices for 1~ms dead time; bottom: difference between the above images. Further details can be found in the main text.}
	\label{fig:rabbit2D}
\end{figure*}

For initial demonstration purposes, we first show \emph{ex-vivo} images of a rabbit head, which was soft boiled in tap water to delay tissue deterioration (Fig.~\ref{fig:rabbit2D}(a)). Figure \ref{fig:rabbit2D}(c) contains selected slices from the full 3D reconstruction employing a PETRA sequence. The field of view is 46x54x42~mm$^3$, with an isotropic resolution of 0.5~mm. We excite the sample with a hard rf pulse of \SI{10}{\micro s} for a flip angle of $\approx$58~degrees, and the repetition time is $\text{TR}=10$~ms. These parameters were used in two independent acquisitions: one with a short dead time (\SI{90}{\micro s}, limited by ring-down in our setup) to read in the combined signal from hard and soft tissues (Fig.~\ref{fig:rabbit2D}(c) top); and one with a long dead time (\SI{1}{ms}) to remove the short-lived contribution of teeth and skull tissues (Fig.~\ref{fig:rabbit2D}(c) middle). Each radial acquisition lasted \SI{710}{\micro s} (\SI{4}{ms}) in the combined (soft) tissue scan, with a total of 4098 (4098) radial spokes, corresponding to an undersampling factor of 7, and 848 (3904) single points to fill the $k$-space gap, requiring 75 (25) averages and a total scan time of 61 (33) minutes for the reconstructions in Fig.~\ref{fig:rabbit2D}(c). Both images are reconstructed with ART using $\lambda = 0.5$ and 2 iterations in Eq.~(\ref{eq:ART}), and denoised with Block-Matched Filters \cite{Maggioni2013}. Finally, we subtract one of the above scans from the other to produce the bottom image in Fig.~\ref{fig:rabbit2D}(c), where only hard tissues are highlighted. Even this basic post-processing is enough to identify the upper (A), bottom (B), and inner (C) teeth, as well as the rabbit jaw (D).

\subsection{Ultra-short $T_2$ tissue imaging: bare human teeth}

\begin{figure}
	\includegraphics[width= 0.9\columnwidth]{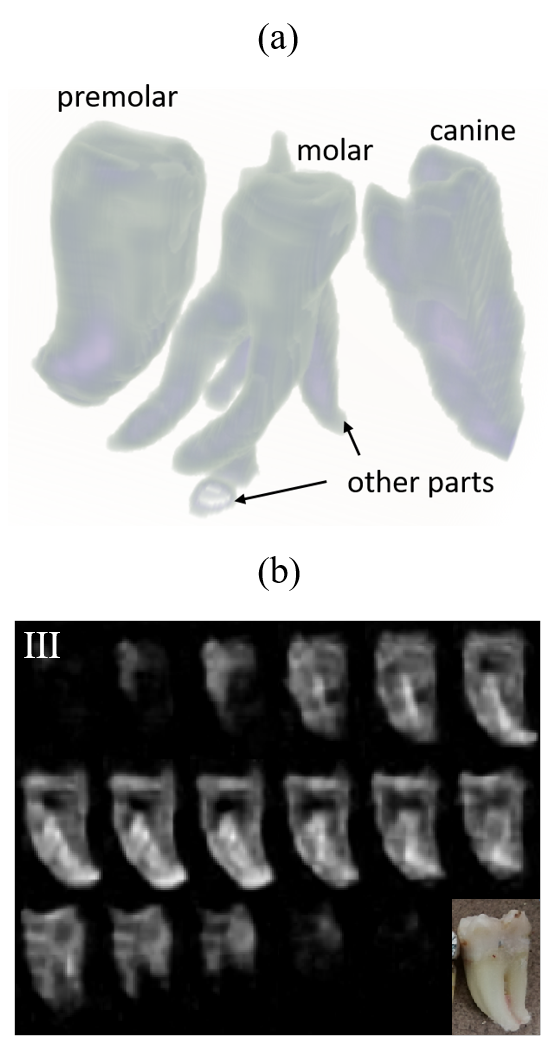}
	\caption{(a) 3 dimensional reconstruction of a set of different human teeth. The image shows a molar, premolar and canine. (b) 2 dimensional slices of the premolar tooth.}
	\label{fig:toothSlices}
\end{figure}

The results shown in Fig.~\ref{fig:rabbit2D} demonstrate the capabilities of ``DentMRI - Gen I'' for imaging of soft and hard tissues simultaneously at sub-Tesla fields. Since human dental structures differ significantly from those in rabbits, below we demonstrate our scanner's performance for a sample consisting of four bare teeth from anonymous donors.

Figure~\ref{fig:toothSlices} shows a 3D renderization of a premolar, a molar and a canine in the sample, and 2D slices of the premolar, obtained from a PETRA scan and applying ART. We acquired the image with a field of view of 42x45x38~mm$^3$ and a nominal isotropic resolution of 0.5~mm. To excite the sample we used a hard pulse of \SI{10}{\micro s} to produce a flip angle of $\approx$90~degrees. The dead time was set to \SI{200}{\micro s} and the acquisition time was \SI{400}{\micro s} with a repetition time of 20~ms. We acquired a total of 3130 radial spokes, corresponding to an undersampling factor of 5, and 7624 single points. To increase the SNR, we averaged over 150 scans with a total scan time of 9 hours. For this ART reconstruction we used $n_\text{it}=1$ with $\lambda=1$.

The samples dried for months before the scan, so there is no pulp present in these pieces and the premolar slices clearly show a dark cavity where there would have been bright soft tissue. Figure~\ref{fig:toothSlices} therefore demonstrates that human teeth can be imaged with high resolution under adverse conditions at low magnetic fields. Obviously, our scan times so far are incompatible with clinical use. We discuss possible approaches to overcome this limitation in Sec.~\ref{sec:concl}.

\subsection{PETRA and DRaNSSE}

%\begin{figure}
%	\includegraphics[width = \columnwidth]{fig_cowbone_image.png}
%	\caption{(a) Photograph of cow bone sample. (b) Raw image slices from 3 dimensional acquisitions with PETRA and DRaNSSE, reconstructed with ART (top row) and FT (bottom). (c) Signal to noise ratio along the red dotted lines in (b). White dashed lines highlight difference between ART and FFT reconstructions.}
%	\label{fig:cowBoneImage}
%\end{figure}
\begin{figure}
\begin{center}
\subfigure[Cow bone sample. ]{\includegraphics[width = 0.5\columnwidth]{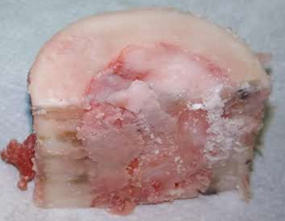}}\hfil
\subfigure[Slices from 3D acquisitions]{\includegraphics[width = \columnwidth]{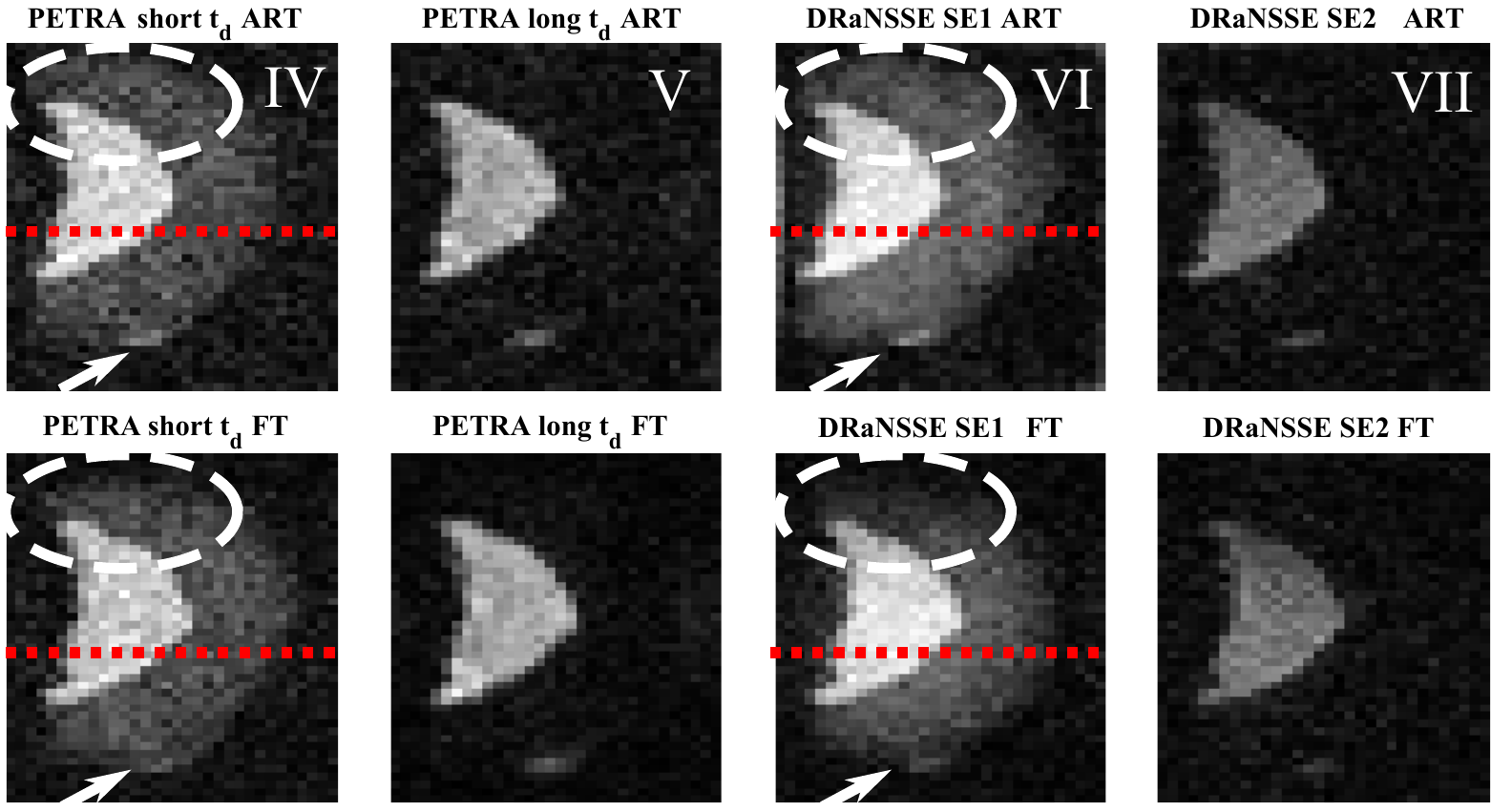}}\hfil
\subfigure[SNR along the red dotted lines]{\includegraphics[width = \columnwidth]{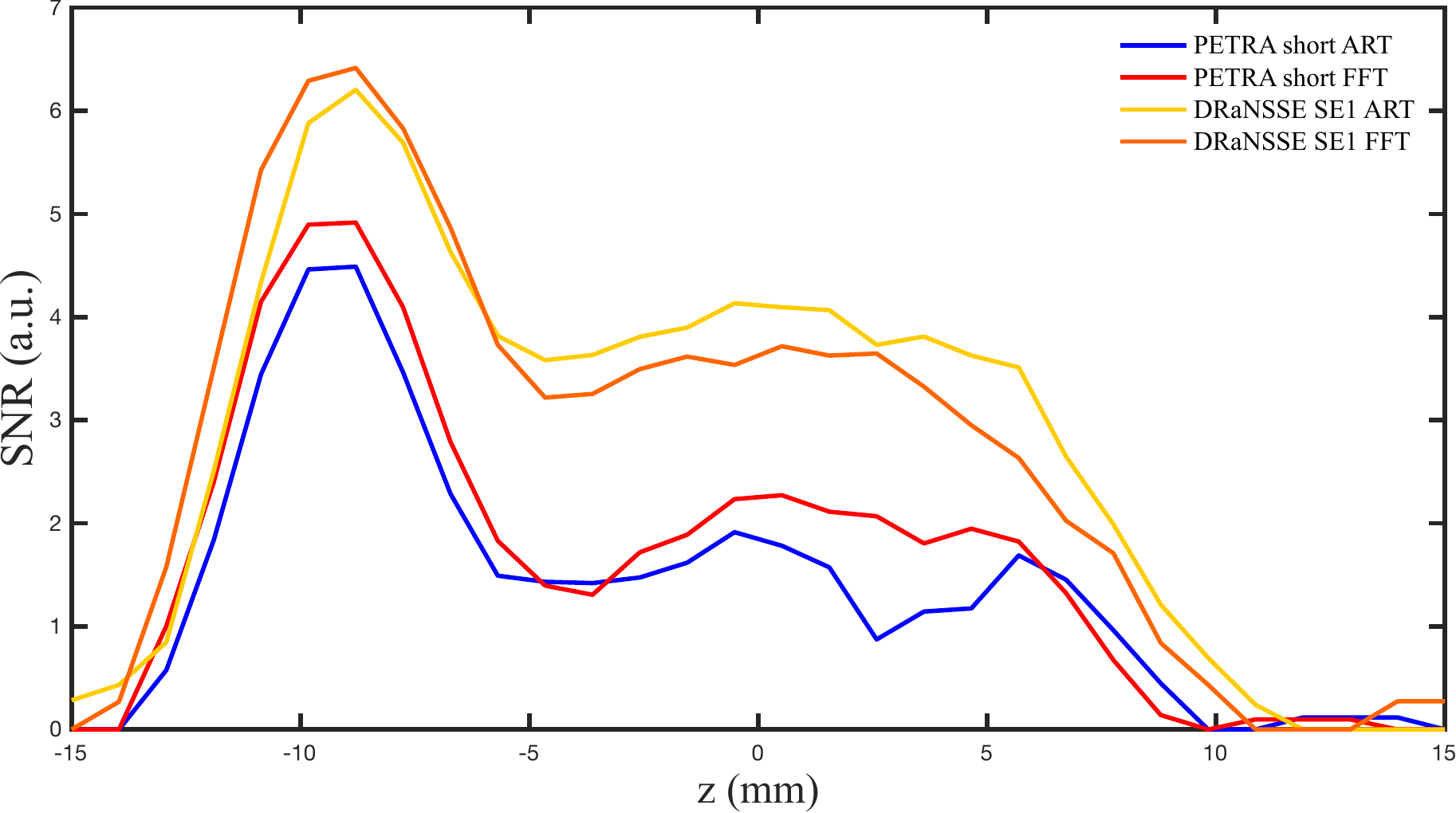}}
\end{center}
	\caption{SNR for different reconstruction methods: (a) Photograph of cow bone sample; (b) Raw image slices from 3 dimensional acquisitions with PETRA and DRaNSSE, reconstructed with ART (top row) and FT (bottom); (c) Signal to noise ratio along the red dotted lines in (b). White dashed lines highlight differences between ART and FT reconstructions.}
	\label{fig:cowBoneImage}
\end{figure}

As shown in Fig.~\ref{fig:rabbit2D}, tissue contrast with $T_2$ weighting with PETRA is possible by image subtraction. This is a lengthy procedure, however, which can be significantly improved with DRaNSSE (Fig.~\ref{fig:seqZTE}(b)). Here we compare one approach against the other.

For these tests we use a piece of a cow femur (Fig.~\ref{fig:cowBoneImage}(a)), which is mostly composed of only two tissues (bone, with $T_2\approx 1$~ms, and marrow, with $T_2 > 50$~ms) and therefore facilitates $T_2$ discrimination and image analysis. In Fig.~\ref{fig:cowBoneImage}(b), images labeled as ``PETRA long (short) $t_\text{d}$'' correspond to an individual PETRA scan with long (short) dead times, where marrow (and bone) appear visible. For DRaNSSE we run a single scan, which can be used to reconstruct marrow (and bone) from the second (first) acquisition in the sequence, corresponding to images with the label ``DRaNSSE SE2 (SE1)''.

Importantly, the total scan times for this study are kept the same, i.e. the sum of both PETRA scan durations is very close to the single DRaNSSE scan time ($\approx 65$~min). Also common to both sequences are: a flip angle of $\approx 90$~degrees, a repetition time of 50~ms, a field of view of 46x48x32~mm$^3$, an isotropic voxel resolution of 1~mm, a sampling rate of 24~kHz (8~kHz for long $t_\text{d}$ PETRA), and a total of 4896 radial spokes in $k$-space, which is fully sampled in this case. For the short (long) $t_\text{d}$ PETRA acquisition we set the dead time to \SI{85}{\micro s} (1~ms), the radial acquisition time to \SI{915}{\micro s} (2~ms), we fill the center of $k$-space with 40 (1528) single points, and we average over 7 (7) acquisitions to increase the SNR with a total scan time of 29 (37) minutes. For DRaNSSE, TE$_1$ is set to \SI{60}{\micro s} and TE$_2$ to \SI{10}{ms}, with a common acquisition time of 1~ms. The overall DRaNSSE scan time was 65 minutes for 16 averages. 

It is apparent from a qualitative comparison between the raw (unfiltered) image sets IV and VI in Fig.~\ref{fig:cowBoneImage}(b), that DRaNSSE reconstructions feature a higher SNR than with PETRA. Figure~\ref{fig:cowBoneImage}(c) shows the SNR  along the red dotted lines in Fig.~\ref{fig:cowBoneImage}(b) for four different cases, quantitatively reinforcing this observation (further details on these calculations can be found in the Appendix). Voxels corresponding to bone and marrow tissue both feature a stronger SNR for DRaNSSE when reconstructed with ART (unlike with Fourier methods, which we discuss below). This is consistent with the fact that signal acquisition in DRaNSSE starts when spins are in full phase coherence after the refocusing pulse, while in PETRA the finite dead time means that spins have already started to dephase before data acquisition. Consequently, for SE$_1$ echo times comparable to the dead time in PETRA, a stronger signal is expected in DRaNSSE. Furthermore, the simultaneous acquisition of both images in DRaNSSE allows for more averaging in the same total scan time. In the scans in Fig.~\ref{fig:cowBoneImage} we acquired 16 averages with DRaNSSE, versus 7 for both short and long dead times with PETRA. In addition to the number of averages, the sampling rate also influences the noise level. We used 24~kHz for images IV, VI and VII and 8~kHz for image V. All in all, the SNR is a factor $\approx1.4$ higher with DRaNSSE than with PETRA for soft tissues, and $\approx2.2$ for hard tissues.

\begin{figure}[b]
	\includegraphics[width = \columnwidth]{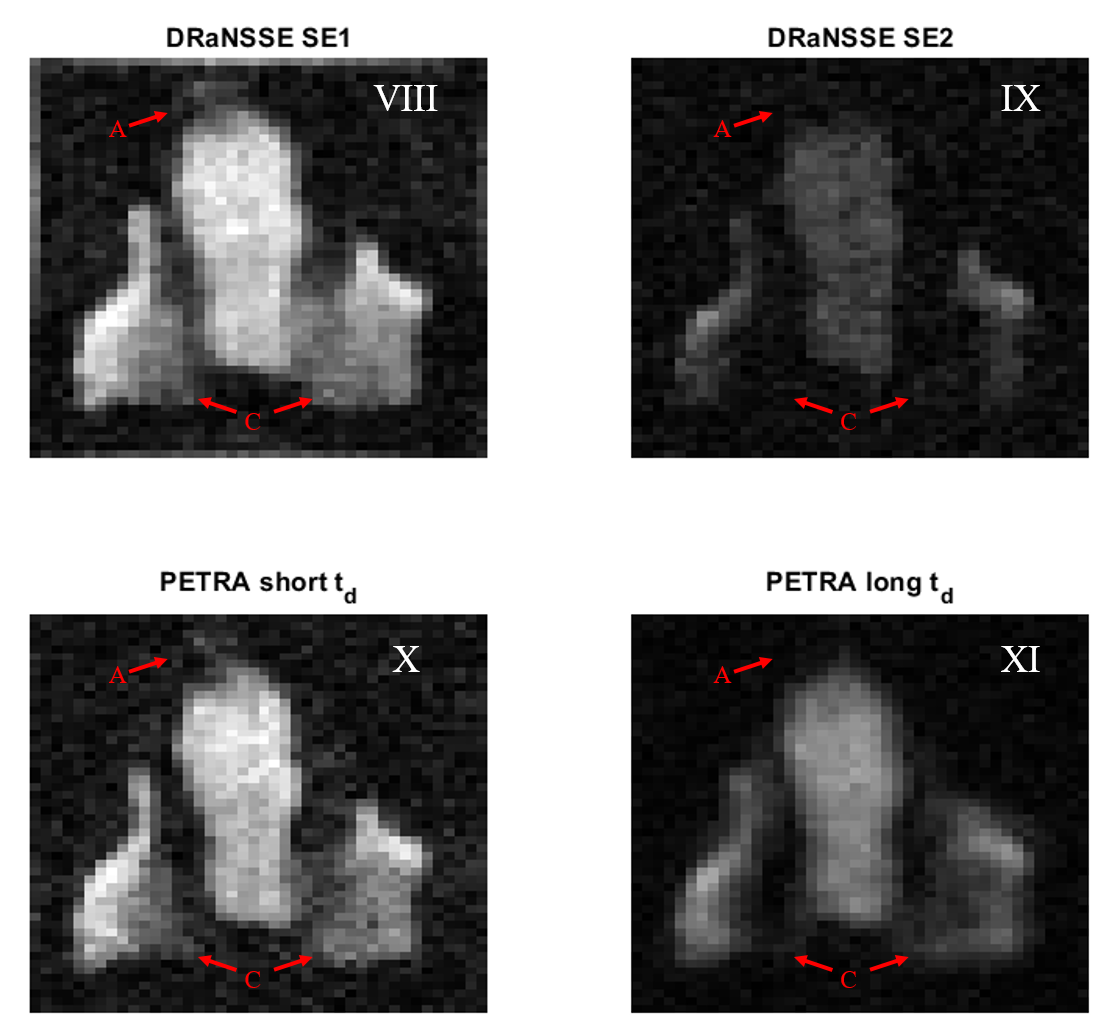}
	\caption{Rabbit image slices obtained with DRaNSSE (top) and PETRA (bottom) for short time parameters (left) and long time parameters (right).}
	\label{fig:Rabbit1mm}
\end{figure}

A similar comparison between DRaNSSE and PETRA acquisitions of a rabbit head shows that tissue contrast is also enhanced with the former. Figure~\ref{fig:Rabbit1mm} shows ART reconstruction slices from DRaNSSE (top) and PETRA (bottom) acquisitions. The right (left) column reconstructions show soft (and hard) tissues. Common to both sequences are: a total scan time of $\approx 30$~min, a flip angle of $\approx 90$~degrees, a repetition time of 50~ms, a field of view of 44x52x42~mm$^3$, an isotropic voxel resolution of 1~mm, a sampling rate of 26~kHz (5.2~kHz for long $t_\text{d}$ PETRA), and a total of 1426 radial spokes in $k$-space, corresponding to an undersampling factor of 5. For the short (long) $t_\text{d}$ PETRA acquisition we set the dead time to \SI{90}{\micro s} (1~ms), the radial acquisition time to \SI{910}{\micro s} (4~ms), we fill the center of $k$-space with 64 (496) single points, and we average over 12 (9) acquisitions to increase the SNR with a total scan time of 15 (15) minutes. For DRaNSSE, TE$_1$ is set to \SI{60}{\micro s} and TE$_2$ to \SI{10}{ms}, with a common acquisition time of 1~ms. The overall DRaNSSE scan time was 31 minutes for 26 averages. All images are produced using $\lambda = 0.3$ and 7 ART iterations. 

The red arrows in Fig.~\ref{fig:Rabbit1mm} show the upper (A) and inner (C) teeth (see Fig.~\ref{fig:rabbit2D}(b)). The dead time of 1~ms in PETRA (image XI) is not long enough to fully remove the hard tissue signal, making it difficult to identify hard tissues from a comparison with X. One possible solution is to increase the dead time, but this requires longer overall scan times. For instance, if we double the dead time to $t_\text{d} = 2$~ms (still significantly shorter than the 10~ms echo time of SE$_2$ in IX), then we need to increase the number of single points at the center of $k$-space by a factor $2^3$. This adds up to 3968 single points keeping the rest of the above settings, prolonging the total scan time to $\approx50$~min. A second possibility is to increase the readout window proportionally to the dead time, i.e. going to an acquisition time of 8~ms in our example. However, such a long acquisition time is affected by $T_2^*$ decay, which degrades the sharpness of the image contours. This effect is already visible in XI for an acquisition time of 4~ms.

A further advantage of DRaNSSE with respect to PETRA is that the relevant decoherence time constant is $T_2$ rather than $T_2^*$, as a result of the induced spin echoes. This means we can wait until hard tissue signals have been strongly suppressed. Consequently, a comparison between images VIII and IX, where SE$_2$=10~ms, shows a starker tissue contrast in regions $A$ and $C$ than between images X and XI.

Finally, a concluding remark regarding DRaNSSE: while it clearly performs better than PETRA in our setup, we occasionally observe reconstruction artifacts originated by stimulated echoes. One possible strategy to overcome this is to tailor the timings so as to not capture them, although it might prove inconvenient. It is therefore relevant to carefully calibrate coherent manipulations in the setup, to minimize the detrimental effect of stimulated echoes.

%Fourier VS ART----------------------

\subsection{Fourier transform vs Algebraic Reconstruction Techniques}

Along with the choice of pulse sequence, the mathematical method employed strongly impacts the reconstruction quality. We used ART in most of the reconstructions shown so far, since we find it performs better than traditional Fourier analysis with our radial $k$-space sampling. Here we look into this topic in more depth.

The results in Fig.~\ref{fig:cowBoneImage}(c) allow for a first comparison. One obvious conclusion is that images IV and VI differ significantly in the encircled regions; due to limited SNR, Fourier techniques artificially darken the area, whereas the signal intensity with ART is more homogeneous. A second visible effect is that the contrast between the small lump of soft tissue and the external surface of the bone where it is attached (white arrows), is higher with ART than with Fourier transformation.

\begin{figure*}[h]
\centering{
\subfigure[Cow bone (1mm) with DRaNSSE]{\includegraphics[width = 0.9\columnwidth]{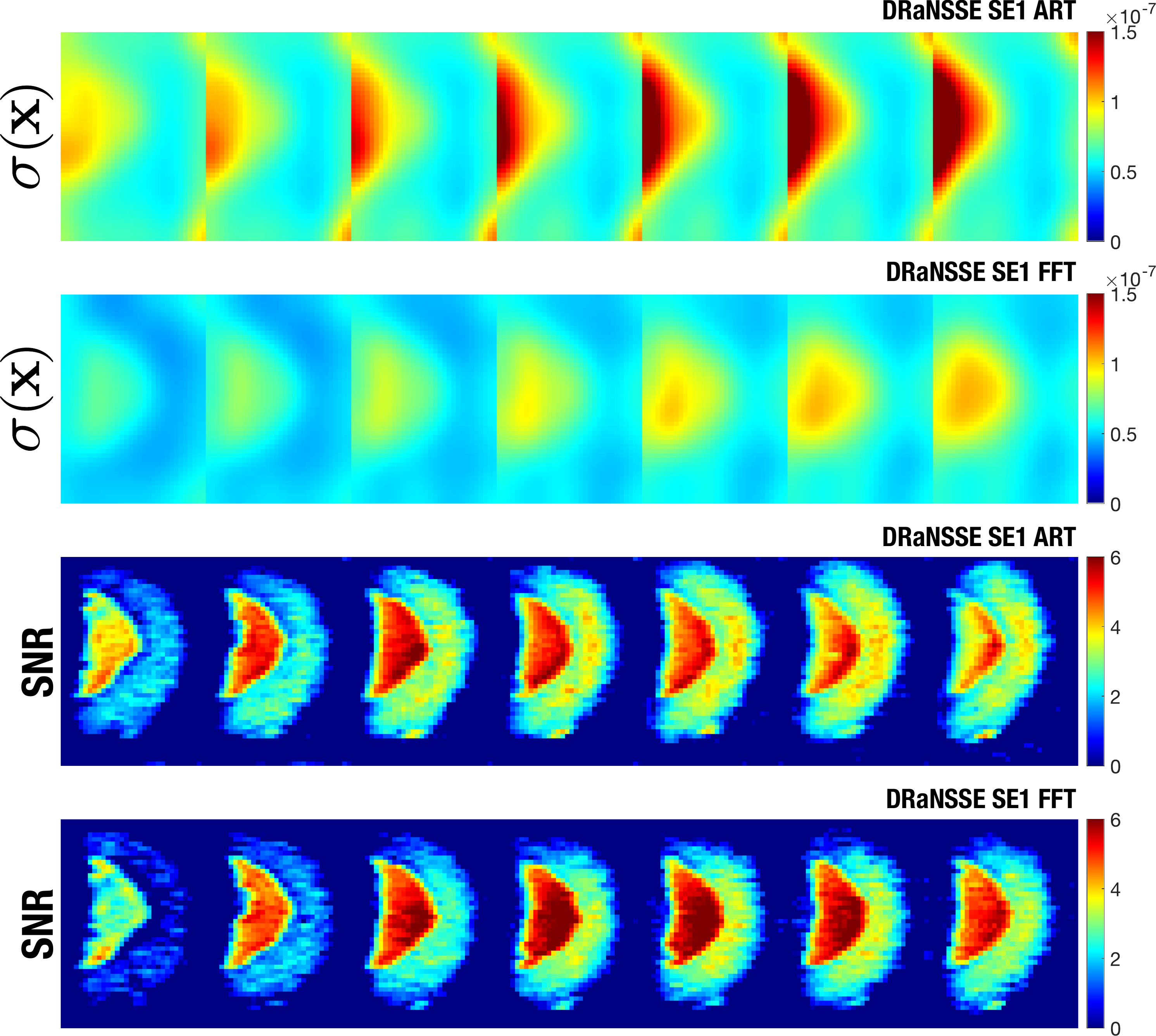}}\hfil
\subfigure[Cow bone (1mm) with PETRA]{\includegraphics[width = 0.9\columnwidth]{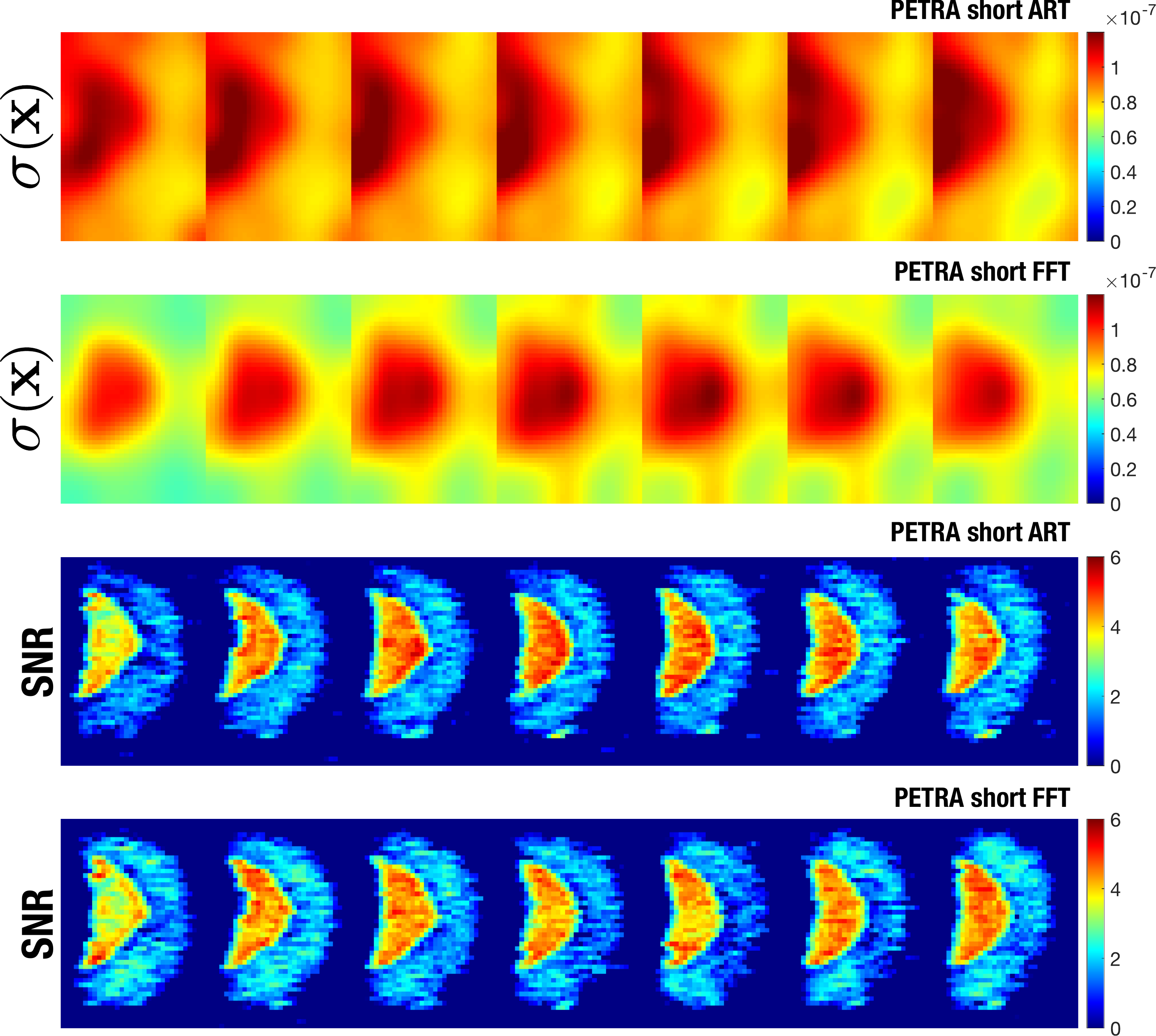}}
%	\caption{Noise and SNR maps for ART and Fourier transform for COW.(a) figure VI, (b) figure V,}
%	\label{fig:noiseANDsnr:cow}
%\end{figure*}
%
%\begin{figure*}[h]
%\centering{
\subfigure[Rabbit bone (1mm) with PETRA]{\includegraphics[width = 0.9\columnwidth]{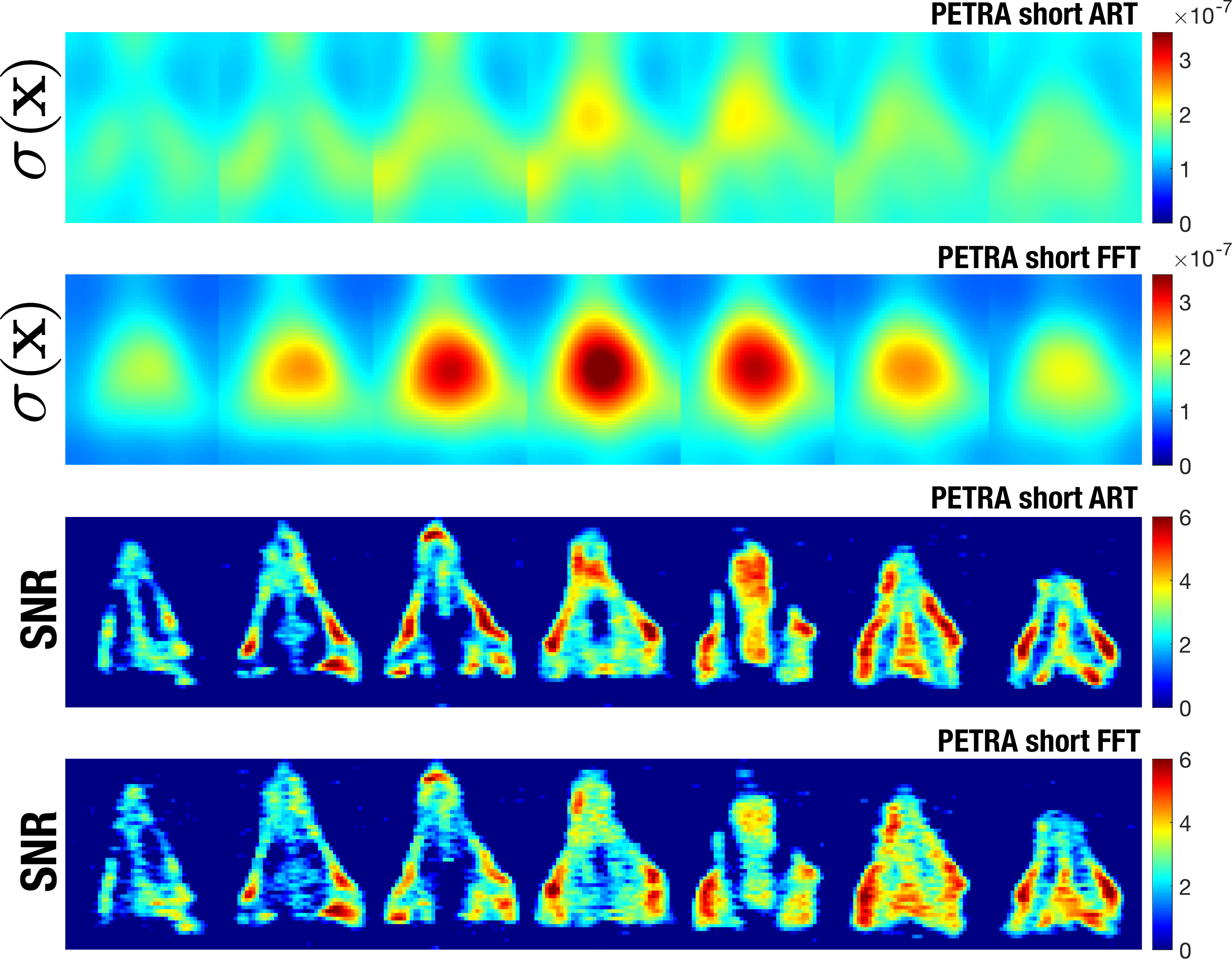}}\hfil
\subfigure[Rabbit bone (05mm) with PETRA]{\includegraphics[width = 0.9\columnwidth]{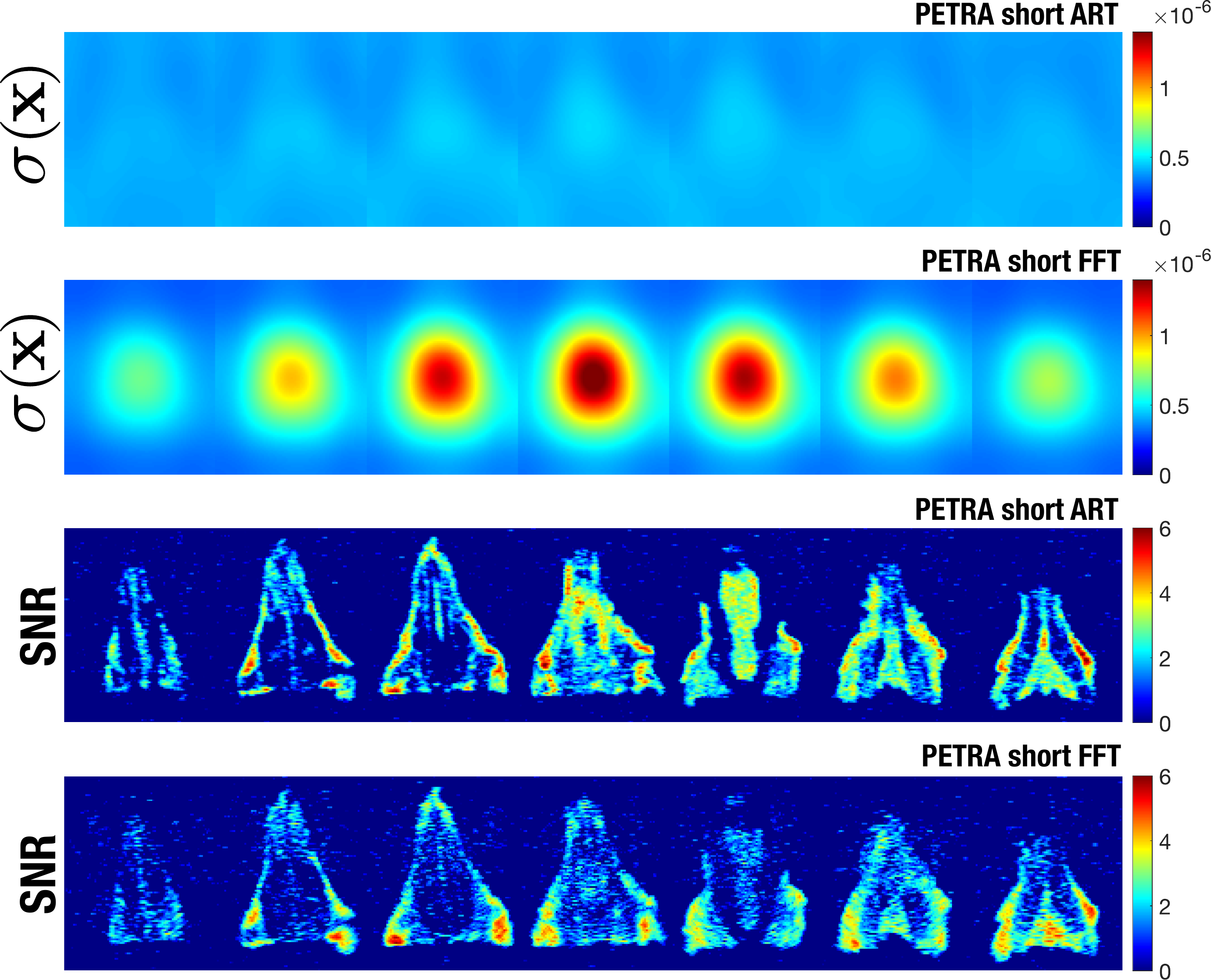}}}
%	\caption{Noise and SNR maps for ART and Fourier transform for RABBIT.(a) figure X, (b) figure I,}
%	\label{fig:noiseANDsnr:rabbit}
	\caption{Noise and SNR maps for ART and Fourier transform for images VI (a), V (b), X (c) and I (d).}
	\label{fig:noiseANDsnr}
\end{figure*}

Figure~\ref{fig:noiseANDsnr} shows noise and SNR maps for ART and Fourier reconstructions of the cow femur (Fig.~\ref{fig:noiseANDsnr}(a) with DRaNSSE and Fig.~\ref{fig:noiseANDsnr}(b) with PETRA) and rabbit images (Fig.~\ref{fig:noiseANDsnr}(c-d) both with PETRA and 1~mm and 0.5~mm, respectively), following the method described in the Appendix. We observe significant differences between the noise pattern obtained from Fourier reconstructions respect to ART. The former present a highly non-stationary noise for all images (i.e. the variance of noise $\sigma({\bf x})$ depends on position). This is reflected by the coefficient variation (CV) of noise, as well as the average noise level, both shown in Table~\ref{tab:noiseANDsnr}. We also observe for Fourier reconstructions that the maximum noise level happens at the gradient isocenter position, which is coherent with the radial sampling scheme due to the high sampling density close to the center of $k$-space. For ART, the matrix size strongly affects the level of non-stationarity, where the noise variation coefficient is 8 times smaller when compared with Fourier reconstruction for Fig.~\ref{fig:noiseANDsnr}(d), with double matrix size. 
%\JM{Can this be related to the SNR calculation method? If yes, we should mention}. 
All in all, we observe that, while PETRA images reconstructed with ART present a more stationary noise than Fourier reconstruction, the noise obtained with DRaNSSE is highly non-stationary even when reconstructed with ART.

\begin{table*}[t]
\caption{Estimated noise and SNR parameters corresponding to images in Fig.~\ref{fig:noiseANDsnr}. SNR in soft and hard tissue on Figs.~\ref{fig:noiseANDsnr}(c-d) correspond to tongue and inner tooth, respectively. CV stands for Coefficient of Variation (the standard deviation divided by the mean) and $\langle .\rangle$ is the average operator.}
\label{tab:noiseANDsnr}
\centering
\begin{tabular}{|l|c|c|c|c|c|c|}
\hline
Data set&Figure    & \begin{tabular}[c]{@{}c@{}}$\langle\sigma({\bf x})\rangle$\\ ART / FT\end{tabular} & \begin{tabular}[c]{@{}c@{}}CV($\sigma({\bf x})$) \\ ART / FT\end{tabular} & \begin{tabular}[c]{@{}c@{}}$\langle\text{SNR}({\bf x})\rangle$\\ 
ART / FFT\end{tabular} & \begin{tabular}[c]{@{}c@{}}SNR soft tissue\\ 
ART / FT\end{tabular} & \begin{tabular}[c]{@{}c@{}}SNR hard tissue\\ ART / FT\end{tabular} \\ \hline
Cow 1~mm DRaNSSE&   Fig.~\ref{fig:noiseANDsnr}(a) & 0.81 / 0.60 ($\times 10^{-7}$)   & 0.35 / 0.23      & 3.2 / 2.8    & 5.7 / 7.2   & 5.0 / 3.9  \\ \hline
Cow 1~mm PETRA&     Fig.~\ref{fig:noiseANDsnr}(b) & 0.92 / 0.74 ($\times 10^{-7}$)   & 0.14 / 0.20      & 2.4 / 2.4    & 4.7 / 4.0   & 2.1 / 1.4  \\ \hline
Rabbit 1~mm PETRA&  Fig.~\ref{fig:noiseANDsnr}(c) & 1.48 / 1.43 ($\times 10^{-7}$)   & 0.16 / 0.37      & 2.9 / 2.8    & 4.9 / 3.8   & 2.3 / 2.5   \\ \hline
Rabbit 0.5~mm PETRA&Fig.~\ref{fig:noiseANDsnr}(d) & 4.10 / 4.84 ($\times 10^{-7}$)   & 0.05 / 0.41      & 2.4 / 1.6    & 2.8 / 2.2   & 2.2 / 2.2   \\ \hline
\end{tabular}
\end{table*}

In addition to the noise performance, there are also important differences in the SNR. Figures~\ref{fig:noiseANDsnr}(a-b) show that ART reconstructions for both DRaNSSE and PETRA produce higher SNR than Fourier reconstructions in the hard tissue. Generally, ART produces SNR values that, in the worst case, are similar to those from Fourier reconstructions (see Fig.~\ref{fig:noiseANDsnr}(c-d) and Table~\ref{tab:noiseANDsnr}). Moreover, note that ART yields better results than Fourier transforms both when $k$-space is fully sampled (Fig.~\ref{fig:noiseANDsnr}(a-b)) and when it is undersampled (Fig.~\ref{fig:noiseANDsnr}(c-d)). Finally, since the results with ART depend on the reconstruction parameters $\lambda$ and $n_\text{it}$, and there is therefore room for potential improvement with respect to the results we present here, we conclude that ART is better suited for image reconstruction than Fourier analysis with our settings.

\section{Conclusion}
\label{sec:concl}

In summary: we have demonstrated the capability of our new low-cost ``DentMRI - Gen I'' scanner to simultaneously image hard and soft biological tissues; we have devised a new pulse sequence (DRaNSSE) that, compared to standard short $T_2^*$ sequences such as PETRA, yields higher SNR images and enhanced tissue contrast; and we have shown that iterative techniques (ART) outperform traditional Fourier methods in all quantified metrics, except for the computational time required for the reconstruction.

Nevertheless, a prospective ``Gen II'' scanner must pose solutions to significant remaining challenges if it is to be compatible with clinical conditions. First and foremost, scan times must be reduced (see last column in Tab.~\ref{tab:imageParameters}). In particular, the human teeth images shown take intolerably long to acquire. Although these were taken under notably adverse conditions (tooth samples were dry and the overall signal levels were artificially low due to lack of the surrounding soft tissues), we do not expect that the ``Gen I'' scanner can produce clinically useful images sufficiently fast. We therefore plan three major upgrades for ``Gen II'': i) we will make use of balanced steady-state free precession protocols, which yield optimal SNR values and can be safely used at low magnetic fields \cite{Carr1958,Sarracanie2015}; ii) we will use quantum dynamical decoupling techniques such as WAHUHA \cite{Waugh1968} or CHASE \cite{Waeber2019} sequences, which can prolong the lifetime of the magnetic resonance signal of hard tissues; and iii) we will perform dual species MRI on protons and $^{31}$P nuclei, since the latter are more abundant in hard biological tissues and they provide complementary information \cite{Frey2012}. Additionally, we are starting to investigate the possibility of slice selection with zero-echo time sequences for fast 2D imaging. Aside from shortening scan times, a reduction in the overall footprint and weight would ease scanner siting. The heaviest (and most expensive) component is the permanent magnet ($\approx$940~kg and $\approx$40~k\euro), which on top imposes the use of a sizable mechanical support structure. For this reason, we are currently working on a more efficient design, and estimate that we can bring the weight down to $\approx$720~kg (and $<10$~k\euro).

Besides dental imaging, a low cost MRI scanner capable of detecting hard tissues and solids may find application in different scenarios, including but not restricted to: head and extremity imaging \cite{Marques2019,Cooley2020}; the food industry (e.g. in inspection and selection tasks, see \cite{Hills2003}); or rubber degradation control, which is relevant in, for example, the industries of mining or transport \cite{Somers2000}.

\section{Contributions}

Experimental data were taken by JMA, EDC and JB, using an apparatus built by DGR, JPR, EDC, JMA, RB, JMG, EP, CG and JA. Image reconstruction and segmentation was performed by JMA, ED, FG, JB and JA. Noise and SNR estimation were performed by SAF. The paper was written by JA and JMA with input from all authors. Experiments conceived by JMB, JA and AR.


%APPENDIX-----------------------------------
%\appendix
\appendices
\section{Estimation of the SNR}
\label{sec:app:snr}

We assume that the final reconstruction signals in image domain follow a non-stationary Rician distribution \cite{AjaBookNoise}, i.e., a Rician distribution where the noise parameter $\sigma$ becomes position dependent: $\sigma({\bf x})$. This assumption is based on two facts:
\begin{enumerate}
\item Gaussianity and Rician: ART and FT reconstructions both produce signals with additive Gaussian Noise in each of the points. The magnitude of that Gaussian data produces Rician data if the real and imaginary parts of the former are independent and have the same $\sigma$ parameter. Although this need not be strictly the case with our reconstruction processes, we show below that the Rician approach yields valid results.
\item Non-stationarity: ART and FT both perform a linear reconstruction of complex Gaussian data where different weights are applied to the original values. As an effect of the weights and the correlations introduced by the reconstruction, the final reconstructed data can show different noise properties in each voxel, i.e., every voxel can show a different value of $\sigma({\bf x})$. In our images, as is usually the case, $\sigma({\bf x})$ varies slowly with ${\bf x}$ and can be considered a low frequency signal \cite{AjaBookNoise,aja2015spatially}.
\end{enumerate}

\begin{figure}[h]
\begin{center}
\includegraphics[width=0.6\columnwidth]{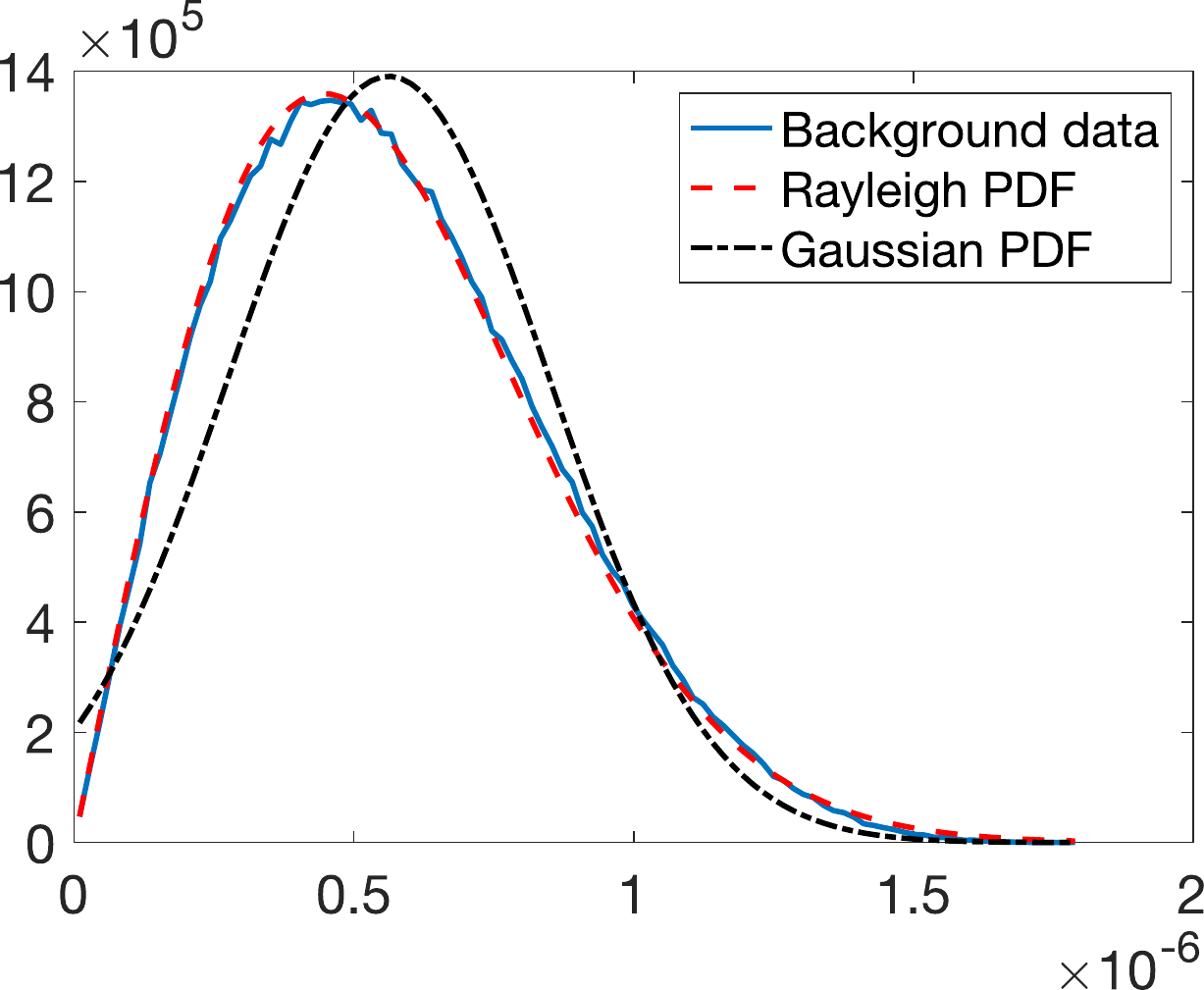}
\end{center}
\caption{Fit of the background data of image I in Fig.~\ref{fig:rabbit2D} to a Rayleigh distribution. A fit to a Normal distribution is shown for comparison.}
\label{fig:app:rayleigh}
\end{figure}

In order to test the first assumption (the Rician nature of the data), we consider Fig.~\ref{fig:noiseANDsnr}(d), where the background noise is to a good approximation stationary and a single value of $\sigma$ characterizes the distribution. The expected signal strength in the background is zero, and the Rician Probability Distribution Function (PDF) simplifies to a Rayleigh distribution. From the fits in Fig.~\ref{fig:app:rayleigh} we conclude that the signal in the background is Rayleigh-distributed, which is compatible with our Rician assumption.

Thus, in order to estimate the SNR of the final volume we assume a non-stationary Rician model. We define the {\em estimated} SNR as
$$
\text{SNR}({\bf x})=\frac{\widehat{A({\bf x})}}{\widehat{\sigma({\bf x})}},
$$
where $\widehat{A({\bf x})}$ and $\widehat{\sigma({\bf x})}$ are, respectively, an estimation of the original signal (in absence of noise) and and estimation of the variance of the noise in each point of the image. 

In order to estimate the noise, we use a 3D extension of the homomorphic scheme proposed in Ref.~\cite{aja2015spatially}. The rough estimation of the original signal is provided by the second order moment of a Rician distribution. If $M(\bf{x})$ is a Rician signal, then \cite{AjaBookNoise}
$$
E\{M^2({\bf x})\}=A^2({\bf x})+2\cdot\sigma^2({\bf x}).
$$
We estimate the expectation $E\{M^2({\bf x})\}$ by a local mean, calculated as the convolution of the original signal with a $3\times 3\times 3$ average kernel, $h({\bf x})$. Hence:
$$
\widehat{A({\bf x})}=\sqrt{\max\{M^2({\bf x})*h({\bf x})-2\cdot\widehat{\sigma^2({\bf x})},0\}},
$$
and the estimated SNR becomes:
$$
\text{SNR}({\bf x})=\frac{\sqrt{\max\{M^2({\bf x})*h({\bf x})-2\cdot\widehat{\sigma^2({\bf x})},0\}}.}{\widehat{\sigma({\bf x})}}.
$$
Note that, due to the use of the convolution kernel, the SNR will show some blurring in the edges of the image. However, for the comparison purposes of this work, that will not affect results.

%\appendices
%% you can choose not to have a title for an appendix
%% if you want by leaving the argument blank
%\section{}

\section*{Acknowledgment}

We thank anonymous donors for their tooth samples, Andrew Webb and Thomas O'Reilly (LUMC) for discussions on hardware and pulse sequences, and Antonio Trist\'an (UVa) for information on reconstruction techniques. This work was supported by the European Commission under Grants 737180 (FET-OPEN: HISTO-MRI) and 481 (ATTRACT: DentMRI). Santiago Aja-Fern\'andez acknowledges Ministerio de Ciencia e Innovaci\'on of Spain for research grant RTI2018-094569-B-I00.

\ifCLASSOPTIONcaptionsoff
  \newpage
\fi

%\bibliography{myrefs}

\end{document}